\documentclass[
 reprint,
 amsmath,amssymb,
 aps, prper
]{revtex4-2}

\usepackage{caption}
\usepackage[font=small,skip=0pt]{caption}
\usepackage[utf8]{inputenc}

\usepackage{xcolor}
\usepackage{dcolumn}
\usepackage{rotating}
\usepackage{tabularx}
\usepackage{rotating,multirow}
\usepackage{array,graphicx}
\usepackage{subcaption}
\usepackage{booktabs}
\usepackage{hyperref}
\hypersetup{colorlinks=true,urlcolor=blue,citecolor=blue,linkcolor=blue}    
\usepackage{xurl}

\usepackage{enumitem}    
\setlist{nosep}

\begin{document}

\preprint{APS/123-QED}

\title{How the physics culture shapes the experiences of undergraduate women physics majors: A 
comparative case study of three physics departments}

  \author{Lisabeth Marie Santana}
  \author{Chandralekha Singh}
  \affiliation{Department of Physics and Astronomy, University of Pittsburgh, Pittsburgh, PA, USA 15260} 

\date{\today}

\begin{abstract}
    This investigation is a comparative case study of the physics department culture at three institutions based upon the points of view of
    undergraduate women majoring in physics. The three studies conducted in the United States include Johnson's 2020 study 
    in a small physics department at a small predominantly White liberal arts college, Santana and Singh's 2023 study at a large predominantly White research institution, and Santana and Singh's 2024 study in a medium-sized physics department at a small predominantly White private liberal arts college.
    Using synergistic frameworks such as Standpoint Theory, Domains of Power, and the Holistic Ecosystem for Learning Physics in an Inclusive and Equitable Environment (HELPIEE), we compare and analyze the narratives of undergraduate women. Their reflections are valuable for understanding how those in the position of power, e.g., instructors, have important roles in establishing and maintaining safe, equitable, and inclusive environments for undergraduate students. Their accounts help us contrast the experiences of undergraduate women in physics departments with very different cultures.
    This comparative analysis focusing on the experiences of undergraduate physics majors in departments with drastically different cultures is especially important for reflecting upon what can be done to improve the physics culture so that historically marginalized students such as women and ethnic and racial minority students in physics feel supported and thrive.
    In particular, this comparative case study can be invaluable to understand the positive and negative aspects of the physics culture as they impact undergraduate women majoring in physics within these three departments. This analysis can help other physics departments contemplate strategies to improve the physics culture so that all undergraduate physics majors 
    have validating experiences while navigating their physics journey regardless of their identities.
\end{abstract}

\maketitle

\section{Introduction}

Physics has been historically notorious for marginalizing certain groups of students, such as women and ethnic and racial minorities (ERM). Reports within the recent decade reveal that only 20-25\% of physics bachelor's degrees in the US have been awarded to women \cite{AIPbachelorsdegrees2018, AIPbachelorsdegrees2020, APS2020PhysicsDegrees}. 
2020 reports also show that of physics bachelor's degrees in the US, 3-4\% are awarded to Black students and 12\% are awarded to Latinx students \cite{AIP2020Physicsdegrees, boatman2022}.
Understanding this underrepresentation and improving not only the number of physics bachelor's degrees awarded to underrepresented groups, but their experiences in undergraduate physics programs has been an area of interest to physics education researchers.

Based on findings from previous studies, some challenges that women and ERM students in physics face relate to their identities. These challenges may partly be due to physics programs not having the representation of women or ERM students that is reflective of society \cite{seymour1997talking, seymour2019talking, pollack2016, hazari2007genderdifferences, sax2016wip, potvin2015underpresentation, hazari2020context, lorenzo20066gendergap, Cheryan2017STEMGender, jones2000gender, hill2010few, ganley2018gender, mccullough2011women}. This lack of accurate representation, or underrepresentation, can have negative impacts on students, such as leading them to question whether or not they belong in physics. Previous findings also reveal that sense of belonging is tied to self efficacy as well as perceived recognition and
women and ERM students often have negative experiences in the physics learning environments \cite{barthelemy2016gender, traxler2016genderinphys, rosa2016obstacles, santana2022investigating,santana2023effects}. We emphasize that these lived experiences are very important in understanding how marginalized students navigate their physics journey. These negative experiences can lead to gaps in performance and psychological factors (e.g., self-efficacy, belonging, identity, etc.) disadvantaging traditionally marginalized students in physics further.

Previous studies have found that amongst men and women students, gaps in performance as well as in their physics psychological characteristics are prevalent throughout high school and college.
Researchers found gender gaps such as women having lower perceived recognition compared to men as a ``physics person" from their peers and instructors \cite{hazari2010connecting, kalender2020damage, kalender2019female, nissen2016gender, zeldin2008comparative, lock2015physidentity, cwik2022not, liimpact2023}. This gap in perceived recognition can negatively affect women's academic performance \cite{cwik2021perception} and influence their decisions to leave the field \cite{sawtelle2012exploring, seymour1997talking, seymour2019talking, good2012optout}.
For example, if women believe that their instructors or peers do not see them as being good at physics, this negative perceived recognition can make them question whether or not they can excel in physics \cite{li2020perception,li2021effect, lock2015physidentity, hazari2015powestructures}, i.e., low or negative perceived recognition may lead to low self-efficacy and vice versa.

Physics is not only notorious for the marginalization of certain student groups, but also for the masculine nature of its social culture \cite{danielsson2012doingphysics}. At many institutions, this type of culture is still prevalent and many physicists still uphold the beliefs that physics is only for smart 
men. Although social culture is difficult to change, and it takes a long time for change to occur, the field of physics education can help 
play an important role in this change in culture. However, a critical number of physics faculty in a given department need to be onboard with this change to catalyze and be sustained.

It is highly concerning that among the natural sciences, physics has the worst stereotypes regarding who belongs in it, who can excel in it, and what a ``traditional" physicist looks like \cite{leslie2015expectations, santos2017you}. For example, based upon stereotypes, a traditional physicist needs to be a genius, thus physics is associated with brilliance which is typically attributed to men\cite{bian2018brilliance}.
In combination with lack of representation, student performance can be negatively affected by these negative stereotypes
\cite{marchand2013stereotypes, cwik2021damage, maries2018agreeing}. 
These two factors may further reinforce each other, especially if physics culture is masculine. For example, if ERM students don't see others like them, or don't have role models who look like them, they are continuously reminded, e.g., that physics is a domain belonging to White men, which can ultimately lead them to not wanting to continue in physics. 
This also portrays an image that one needs to be of a certain demographic group, or should be a genius in order to belong and succeed in physics \cite{pollack2016, allen2016women, maries2018agreeing, maries2020active, karim2018evidence, walton2015two, steele2010stereotypes, reuben2014stereotypes,  marchand2013stereotypes, kelly2016gender, mccullough2007gender, Gonsalves2016Masculinities}. 
This masculine culture continues to harm women and ERM students such that they may become isolated from lack of role models and a community that could provide support to them \cite{francis2017construction, danielsson2012exploring, Gonsalves2016Masculinities, dennehy2017mentors}.

Prior research also shows that women drop out of STEM disciplines with significantly higher overall grade point average (GPA) than men \cite{maries2022gender}. Thus, we can rule out GPA as being an issue for women students to continue in the STEM majors. There has to be other underlying reasons for them leaving.
The unsupportive physics culture that does not take into account students' lived experiences exacerbates the negative impacts of stereotypes and biases about who belongs in physics as well as lack of role models.
From prior studies, sense of belonging and self-efficacy in physics are closely intertwined \cite{kalender2019gendered, lewis2016fittingin, doucette2020hermione}. To improve a psychological factor such as self-efficacy of students in physics courses, which is a multifaceted construct \cite{bandura1999self}, instructors need to create an equitable and inclusive learning environment. Transforming a physics learning environment is particularly likely to increase underrepresented students' sense of belonging which can in turn increase their self-efficacy and improve their retention \cite{masika2016building, goodenow1993classroom, marshman2018female, marshman2018longitudinal, raelin2014gendered, atwood2010ac,whitcomb2020comparison, felder1995longitudinal}. For example, social psychological interventions are small-scale transformations within classrooms that may be a good first step to combat masculine culture of STEM disciplines\cite{walton2015two}.

In order to create this safe and inclusive environment for students, physics instructors must be committed to these issues. Instructors have the power both within the classroom and beyond it. For example, instructors are responsible for setting the tone of what is acceptable behavior for students, how students should be treated, and establishing classroom norms for peer interactions.
Another important aspect to emphasize is that in many physics programs, men make up the majority for both physics majors and physics faculty. Hence, they have both power and influence to make change.
Thus, it is important to keep these factors in mind as physics departments work to understand and improve the experiences of underrepresented students in physics, such as women and ERM students. It is the traditionally marginalized students who can shed light on how they navigate in the existing physics culture of their departments, their interactions with others in physics, whether or not they feel supported, and how the physics culture impacts them.

Qualitative comparative case studies that focus on undergraduate women, such as the one we present in this paper, can be valuable to learn about the physics culture and whether learning environments are equitable or not directly from students who are from traditionally marginalized groups in physics.
Through their narratives,
we can understand undergraduate women's perspectives in their own voices, listening to their experiences in their own individual contexts and reflecting upon them to improve the physics culture.
Previous studies in a variety of contexts in physics education research have showed the impact and usefulness of student voices.
For example, some prior studies using interviews include those focusing on graduate women experiencing sexism and microaggressions in their programs \cite{barthelemy2016gender}, women in undergraduate labs discussing dynamics of working with male partners \cite{doucette2020hermione}, how students of color are negatively affected by stereotypes in physics \cite{rosa2016obstacles}, 
graduate women of color negatively impacted by stereotypes \cite{santana2022investigating}, and undergraduate women experiencing a negative masculine physics environment \cite{santana2023effects}.  Work by Gonsalves and Danielsson has focused on various issues involving gender in physics, e.g., how women do gender vs. do physics and what traits (feminine vs. masculine) are associated with physics competence 
\cite{danielsson2012exploring, danielsson2014gender, gonsalves2014physics, Gonsalves2016Masculinities}.
Describing a physics environment as masculine, their work demonstrates characteristics that are associated with men, such as 
physics having a competitive culture, physicists exhibiting condescending behaviors, and pervasiveness of stereotypes regarding who can do physics and what a stereotypical physicist looks like.
These studies reveal common stories that show that women and women of color experience hostile environments within physics departments.

Educational institutions, such as universities and colleges, vary in size and the physics departments at different types of institutions also vary in size. The size of a physics department is one factor that can impact how easy or challenging it is for a department to cultivate a physics culture which is different from the prototypical culture\cite{johnson2020intersectional} and whether students from traditionally marginalized backgrounds such as women attending these institutions feel supported. 
In particular, it is important to understand how students who do not identify with those who are dominant in physics, such as women,
navigate through the physics culture especially in departments of different types and sizes, such as large research institutions vs small teaching-focused colleges.
We also note that researchers such as Kanim and Cid have called for physics education researchers to design studies that focus on different types of institutions to create a better representation of educational research to benefit the broader physics community \cite{kanim2020demographics}. They have pointed out that the findings of physics education research may be very different for different types of institutions \cite{kanim2020demographics}. This study focuses on comparative case study concentrating on the physics culture primarily narrated by undergraduate women in physics departments of different types and sizes.
We note that in the literature, apart from Johnson's study \cite{johnson2020intersectional} in a small physics department at a small liberal arts college, and Santana and Singh's study \cite{santana2024importance} in medium-sized physics department at a small liberal arts college, there are little to no other qualitative studies that focus on the experiences of women physics majors in physics learning environments at small liberal arts colleges with physics departments of any size in the US.
Furthermore, we chose another one of our studies which is the only qualitative study we are aware of focusing on the experiences of undergraduate women majoring in physics \cite{santana2023effects} at a large research university in the US.

Thus, this paper focuses on comparative case study, in which we compare three qualitative studies to understand the physics culture primarily as perceived by the women physics majors in those departments in the US. The first study is Johnson's study in which she investigated elements of a highly supportive undergraduate physics program that support students, especially women of color \cite{johnson2020intersectional}. We will refer to Johnson's study as Study 1. This study is of great interest because of the positive elements 
that other physics departments should try to emulate.

The second study, conducted by us, revealed a masculine physics culture that did not support undergraduate women \cite{santana2023effects}. We chose this study for this comparative analysis as 
women physics majors described it to have an opposite environment (not equitable or inclusive) to that described in Johnson's study. We will refer to this study as Study 2.
The third study that we use for this comparative analysis is one we conducted which shows a physics culture somewhere between those manifested in Study 1 and Study 2 as narrated by undergraduate women in physics and shows how women with a variety of intersectional identities navigate a medium-sized undergraduate physics program at a small liberal arts college \cite{santana2024importance}. The women in this study had mixed experiences. We refer to this study as Study 3. 

In Ref \cite{santana2024importance}, we provided a very brief summary of the comparison of these studies in the broader discussion section using the Domains of Power framework.
In this work, we elaborate on that and provide a more in-depth comparison of the physics culture of three physics departments in the US based upon undergraduate women's reflections in these three studies
primarily using the Domains of Power but also taking inspiration from other synergistic frameworks.
The goal of this comparative case study is to highlight the differences and similarities amongst these three US institutions of different types with regard to their physics cultures primarily using undergraduate women's accounts of their experiences navigating the physics learning environments (although we take advantage of interviews with faculty as additional evidence to support the accounts of the physics culture narrated by undergraduate women in Study 1). This comparative case study can help physics departments contemplate how to improve their own physics culture. There are important positive and negative elements described in studies 1-3 which have important implications for any physics department.

\section{Methodology used in the three Studies}
Before we describe the methodology for comparative case study discussed here, we first summarize the frameworks and methodology used for each of the three studies to set the context. We note that all three institutions are predominantly White institutions, meaning people of color are additionally underrepresented in their programs.

\subsection{Frameworks}
We first summarize the three frameworks 
used in the three studies. The three frameworks are Standpoint theory, the Domains of Power, and the Holistic Ecosystem for Learning Physics in an Inclusive and Equitable Environment (HELPIEE).

\subsubsection{Standpoint Theory}

Standpoint theory is a critical theory that focuses on the relationship between the production of knowledge and acts of power \cite{longino1993feminist,harding2004feminist, rolin2009standpoint}. It is related to other critical theories in that it centers around the standpoint or voices of the  underrepresented groups that do not have the same privilege as the dominant group in order to gain a clearer understanding of their struggles.
In this framework, the emphasis is placed on the experiences of undergraduate women in physics to understand what physics departments and instructors can do to improve the physics culture such that they feel safe and supported \cite{blickenstaff2005leakypipeline}.
The nature of all these qualitative studies that we compare here to gain deeper insights into the variety of physics cultures in those physics departments aligns well with Standpoint theory. 
It is through Standpoint theory, that we gain insight to how women perceive and experience their respective physics environment. Therefore, using their stories, we gain information about the people and settings they interact with. Their stories also reveal how power is organized in a physics department.

\subsubsection{Domains of Power}

\begin{figure}[tb]
\small
\begin{center}
\includegraphics[width=0.4\textwidth]{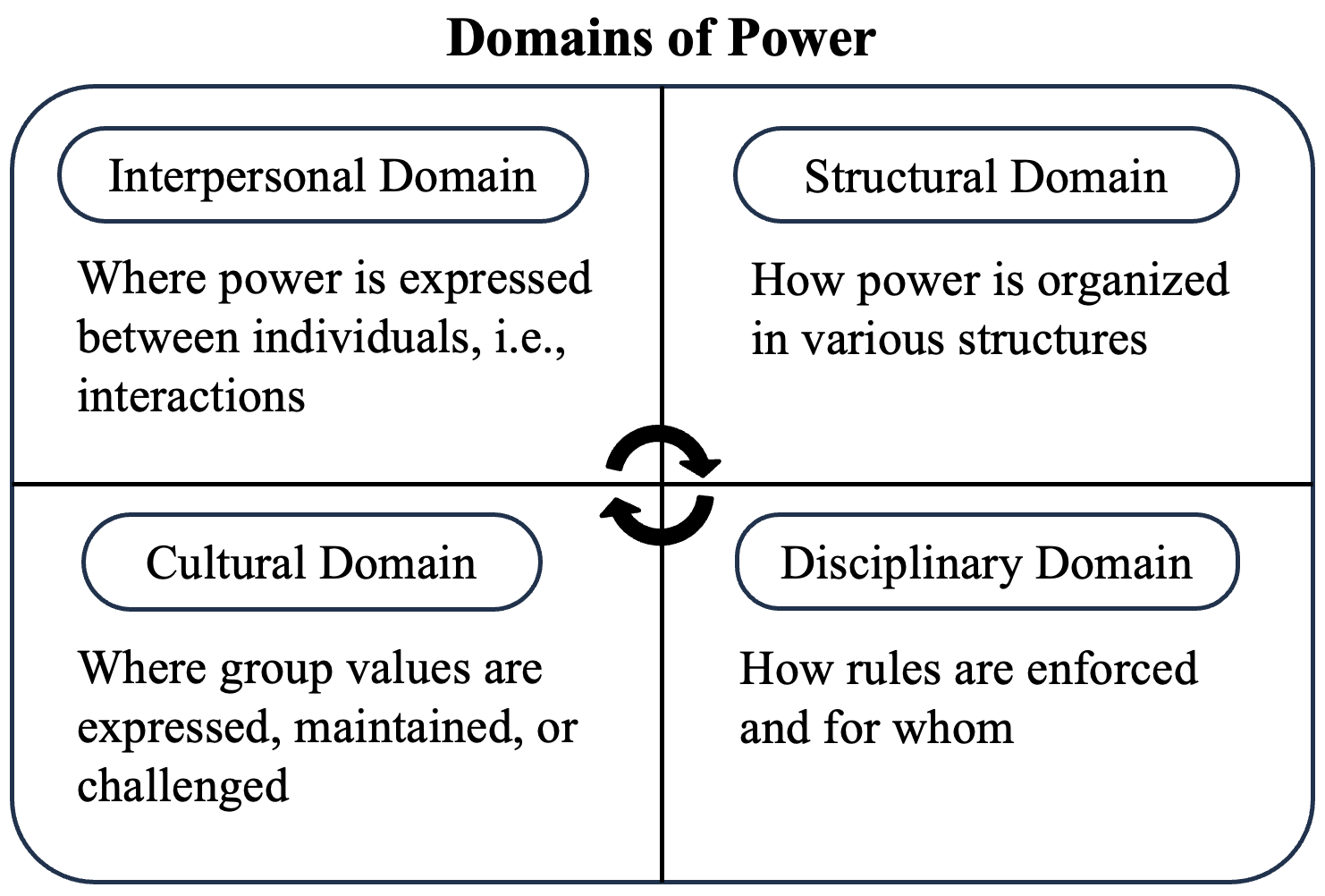}
\end{center}
\caption{The four domains of power: interpersonal, cultural, structural, and disciplinary.}\label{fig:dop}
\end{figure}

Collins introduced four domains as necessary to understand how power is organized in a particular context \cite{collins2009another}. These contexts are essential for understanding what the setting says with regard to who belongs or who has opportunities. She argued that the four domains we need to consider are: interpersonal, cultural, structural, and disciplinary. The interpersonal domain focuses on where power is expressed between individuals. The cultural domain focuses on where group values are expressed, maintained or challenged. The structural domain focuses on how power is organized in various structures.
The disciplinary domain focuses on how rules are enforced and for whom.
Figure \ref{fig:dop} illustrates the four domains.
Within the Domains of Power framework, it is the instructors who have the power over shaping the narrative for each of these domains (which interact with each other). 

Johnson \cite{johnson2020intersectional} noted that in the context of student experiences in learning physics, the interpersonal domain refers to how students communicate with each other and how students and faculty interact; the cultural domain refers to the physics culture; the structural domain refers to the structure of a physics learning environment; and the disciplinary domain refers to how instructors discipline students in the physics courses and in the physics learning environments in general, if their conduct does not conform to the expected norms. Table \ref{table:4.1} is an excerpt from Study 1, showing how the Domains of Power are applied to a prototypical physics department. As a concrete example, in Study 1 \cite{johnson2020intersectional}, the structural domain was analyzed on a small scale focusing on classroom structure.

\begin{table*}[tb]
\small
\begin{center}
\begin{minipage}{0.95\textwidth}
    \caption{\raggedright This table is adapted from Johnson's study \cite{johnson2020intersectional}. The Domains of Power Framework in prototypical physics departments.}\label{table:4.1}
    \begin{ruledtabular}
    \begin{tabular}{ll}
    Domain & Typical physics departments \\ \hline
    Interpersonal & Women of color experience isolation and experience microaggressions. \\ 
    Cultural & Physics is a competition; good physicists are natural geniuses who work in isolation. \\
    Structural & Classes consist of faculty lecturing; high levels of student misconception and failure are expected. \\
    Disciplining & Faculty do not intervene in student-student interactions. \\    
    \end{tabular}
    \end{ruledtabular}
\end{minipage}
\end{center}
\end{table*}

\subsubsection{HELPIEE}
The Holistic Ecosystem for Learning Physics in an Inclusive and Equitable Environment (HELPIEE) framework posits that those in the position of power, e.g., instructors in physics classrooms, have the power to help all students feel supported by carefully taking into account students' characteristics and implementing effective pedagogies in an equitable and inclusive learning environment \cite{cwik2023framework}. 
Within this framework, if students from different demographic groups in a course do not have similar positive experiences and feelings of being supported, the learning environments are not equitable because those in the position of power did not provide adequate support to level the playing field.

Both the HELPIEE and Domains of Power frameworks are synergistic. From these two frameworks, we see that in the context of physics courses, it is the department and instructors' responsibility to create an equitable learning environment. Together, these synergistic frameworks are useful for analyzing these studies with regard to their prevailing physics culture and comparing them.

\subsection{Study 1}
In 2020 \cite{johnson2020intersectional}, Johnson conducted open-ended interviews at a small liberal arts college with 6 undergraduate women majoring in physics, 4 physics faculty, and a focus group consisting of third year physics students. As noted earlier, although our focus here in the comparative case study is primarily on undergraduate women's voices, we take advantage of interviews with faculty as additional evidence to support the accounts of the physics culture narrated by undergraduate women.
The student interviews lasted for about an hour. 
She asked undergraduate women during the individual interviews open-ended questions like ``Tell me your life story in physics."  Out of the individual interviews with undergraduate women, three were women of color and three were White women. At this institution, 25\% of physics majors are women. 
For focus groups, she asked students about their trajectory in physics, what students like about their physics classes, what could improve, and what ideas they might have for attracting more women to physics. 
In faculty interviews, she asked questions such as: ``What does it mean for someone to be a good physics student? What do you do as an individual to support and teach students? What has it been like for you to be faculty at this institution?".

Johnson used the Domains of Power framework to understand intersectional physics identities of students.
She sought to identify and understand the characteristics of a physics department such that women of color felt included and successful. 

\subsection{Study 2}
In Study 2, Santana and Singh conducted semi-structured, empathetic interviews with 16 undergraduate women physics and astronomy majors at a large research university. At the time when the interviews were conducted, this number of women interviewed represents approximately 60\% of the women students in the physics and astronomy department (23\% women). Six out of 16 interviews were analyzed due to the range of their experiences.
At this institution, there are about 50 full-time faculty members.

Each interview was an hour long in duration. The interviews followed protocols set by the researchers prior to conducting the interviews \cite{kvale2009interviews}. Some of the main overarching protocol questions were about students' high school experiences in physics, experiences with college peers and instructors, who or what supported them, and suggestions that might help improve their experiences.
These interviews utilized Standpoint theory to highlight undergraduate women's experiences in physics. 
The goal was to investigate the physics culture in this department based upon undergraduate women major's accounts.

\subsection{Study 3}
In Study 3, Santana and Singh conducted a study similar to their previous study (Study 2), i.e., semi-structured, empathetic interviews with 7 undergraduate women physics and astronomy majors but at a medium-sized
physics department at a small liberal arts college in the US. At this college, women are underrepresented in physics, but not to the extent as they are underrepresented at the national level \cite{AIPbachelorsdegrees2018}, i.e., they make up about one-third of the physics majors.
At this institution, there were 9 full-time faculty members and 4 visiting/adjunct faculty members in this physics and astronomy department. The number of physics majors that graduate each year varies from 10-20.

In order to capture a wide range of experiences, we interviewed women who were at various points in their physics trajectory in college, e.g., second to fourth year students. 
Each interview was about an hour long in duration and was recorded via Zoom. The interviews followed the same protocols as Study 2 set by the researchers prior to conducting the interviews \cite{santana2024importance}. 
The women in our study volunteered to participate in the interviews.
The Domains of Power and HELPIEE frameworks were used to analyze instructors' role in establishing an equitable and inclusive learning environment based upon the accounts of undergraduate women physics majors interviewed.

\section{Positionality}
We now summarize aspects of each researcher's identity that may 
have impacted how they conducted the interviews and analyzed data from their studies.
In Study 1, Johnson is an education researcher, and is a faculty member in a teacher education program. She is a former high school physics teacher. In Studies 2 and 3, both authors are physics education researchers. Santana identifies as a queer Chicana woman. Singh identifies as an Asian-American woman. Together the three researchers have a wide range of experiences in academia. It is important to note these identities as they are also intersectional and were useful when coding and analyzing the intersectional identities of some of these undergraduate women in physics. All researchers have had several prior experiences conducting and analyzing qualitative studies focusing on equity and inclusion in physics learning environments.

\section{Methodology for This Comparative Case Study}

For this comparative case study, we primarily use the Domains of Power framework  
 as a lens for comparing the three studies but Standpoint theory and HELPIEE frameworks, which are synergistic were also valuable. 
Study 1 and Study 3 both already used the Domains of Power framework in the analyses 
so we did not need to re-analyze the data with this lens again. 
However, because Study 2 was only analyzed using Standpoint Theory, we began by discussing, re-analyzing and classifying data presented in Study 2 using the Domains of Power framework.

To begin this re-analysis, we discussed how the interview data from Study 2 fell in each of these domains within the Domains of Power framework (see Figure \ref{fig:dop}).
We note that one of the overarching Analytic Themes in Study 2 was focused on women's interactions with their peers and instructors. These accounts from undergraduate women majoring in physics not only informed us about how power in the physics department in Study 2 is organized in the interpersonal domain but also about the other domains as the four domains of power can be intertwined.
Other big themes such as the Suggestions to Improve Undergraduate Women's Experiences in Study 2 also inform us not only about the disciplinary domain, but the other domains as well. Therefore, the researchers had multiple discussions about these issues between us as we tried to disentangle the domains captured in the quotes and converge on the domain of power that was represented best by a given quote. After multiple discussions, the researchers converged on the classifications of quotes into different domains of power (although since there are overlaps between the four domains, various quotes can potentially be classified as relevant for multiple domains).

Then, we continued to organize the quotes from each study to describe the four domains of power and what these data from the three studies tell us about the overall physics culture. 
We note again that although in this comparative case study, our primary data are undergraduate women's accounts in the three studies, in Study 1, we took advantage of interviews with faculty as additional evidence to support the accounts of the physics culture narrated by undergraduate women.

Thus, in this comparative case study, in the findings described below, the data from each institution are organized in the different domains of power.
After presenting the data from the three studies according to the domains of power, we situate all three studies on a spectrum in the Discussion section.

\section{Findings in each domain}
Here we provide some
quotes from each study within each domain of the Domains of Power  to illustrate the overall physics culture in each department. Since Study 1 was the only study to also conduct interviews with faculty,
we will use those data to supplement the accounts of undergraduate women majoring in physics. 

\subsection{Interpersonal domain}

\subsubsection{Student-student interactions}

In Johnson's study, a second year student summarized her interactions with other physics students:
``Sometimes I'll be in [the physics building]. So I can just like ask a question. Maybe not my class, but it could be an older physics student that could help me. They're super nice! [Q: just women? or women and men?] Either one. Whoever's there. I think I could ask any of them. We're kind of all in it together, why wouldn't they help me? They know what I'm going through – they've done it themselves!''
From this quote, it is clear that students can seek out other physics students, regardless of whether they are in their cohort or not, to ask for help. It seems like there is this camaraderie amongst all physics students, ``We're kind of all in it together" because ``They know what I'm going through." This peer support can have a large impact, as the peer community is welcoming. Students have a good understanding of what the other members of the cohort are experiencing, so they can relate to them.
From this study, we see that the physics culture in this department is such that women have positive interactions with other physics peers regardless of their gender, describing them as friendly, and super nice.

In Study 2, there are many accounts of women having negative experiences with their male peers in different contexts. For example, a senior student describes a male peer as condescending and not someone she would have chosen to work with because he would ignore her inputs and "mansplain" concepts to her (mansplaining is a term that describes men explaining something to a woman in a condescending or overconfident way \cite{mainsplainingdef, solnit2014men, bridges2017gendering}).
She says, ``I would say something and he'd ignore it and I would end up being right and he wouldn't acknowledge that...he ignored me, or I would say something and he would go, \textit{No, no, no, no,} and then he would...explain it to me, like the same thing that I said, like in different words.''
From this quote, it seems like this male student is not receptive towards this student's input. In this physics culture, there seems to be a lack of acknowledgement of this woman's contribution towards the physics problem-solving process.

On the other hand, in Study 2, female students view their female peers as sources of support. Another senior student states:
``...my saving grace that... gave me like confidence and, like the resources that I needed to-to do well in the major... overall [is] my group of friends that I have. I have such an amazing group of four girls that we've all been in the same class together... [those] people are like my rocks and, like my support.''
From this quote we can gather that women lift each other up and form support systems. The culture created by these women seems to 
contrast with the one which is created by male students. Thus, there is a clear dichotomy between male and female students in this regard.

In Study 3, we note that many women described having preferences towards working with female peers, or occasionally male peers only if the two were friends.
For example, one student explains that she mostly works on homework with other female students because of the negative experiences she has had with male peers: 
She says: ``I would generally choose a woman just because...I remember that last semester we would, in one of my physics classes, if me and my partner had different answers, it kind of felt like he'd automatically assume that he was right, which wasn't always the case, and to me that just felt like a very male thing to [do]- So I guess I do generally would select to study with a woman."
Again, through this quote, we get a sense that male peers have this aversion or lack of reception toward their female peers' input during the problem-solving process, similar to Study 2. This student marks this behavior, of ``automatically'' assuming that he was right as a ``very male thing to [do].'' We can infer that in the type of physics culture manifested in this department, female students recognize this masculine behavior and choose to not surround themselves with it.

Moreover, in Study 3, there were several quotes from a woman of color regarding how she felt perceived negatively by her peers. She claims that it is not just men who perceive her this way, but also other women in her physics courses.
She says, ``even amongst a group of other women, there's also been this expectation that I am just not up to par with them, because [of] my racial background and so I feel there's some intersections in terms of my experiences in STEM with women...I definitely know that beyond the realm of just gender, race also plays a really big role in it.''
From this student's perspective, there might not be a dichotomy between White male and female peers because she feels judged by both groups. We also note that this type of physics culture can be extremely difficult to navigate for a student who is one of the few people of color in her program and constantly thinks about how others are perceiving her.

\subsubsection{Student-faculty interactions}

Several women in Study 1 reported that the faculty are accessible and used words such as ``nice" and ``helpful" to describe them. It is important to note that a student reported that both their male and female faculty are accessible. 
One student said that their research advisor ``is like the nicest professor I've ever met in my life... if you do something wrong, she doesn't even frown at you. She's so helpful...everyone loves her...I've heard the general physics kids really like her, and they're the ones who are forced into taking physics and don't actually want to! It seems like she's found a way to really explain things well and get people to be productive without being mean or making you feel bad."
From this student's account, it seems like the physics culture is such that faculty are kind when students make mistakes and help them. In this student's view, this faculty member is good at explaining concepts and does a good job at engaging students, especially those in general physics (i.e., the students who need physics as a requirement such as life science majors).
Not only did students feel comfortable talking to their professors, they explicitly said that they did not feel any more ``intimated by the male professors more than the female professors.''
Johnson also remarked that ``this accessibility is not a coincidence; the faculty make a deliberate choice'' \cite{johnson2020intersectional}.  From faculty interviews that Johnson conducted, one faculty member said: ``I try and make myself really open to if they have questions – just trying to be around the department, so they can find me and ask me if they have questions." 
We see that the physics culture in this department is such that the faculty are making an active attempt to be accessible to students, e.g., by being in their office so that students can drop by and ask them questions.

In Study 2, there are many quotes from undergraduate women about how their male instructors created and fostered negative environments in several setting, similar to male peers. One quote was about a male instructor enforcing negative stereotypes about women in the classroom by calling out a group of women who did not complete the assigned reading.
One student recalled: ``[My instructor] called on one of [the women] and she didn't know the answer. And he was like, \textit{Did you read the book?} And she said, \textit{No, I haven't read it yet}. And she said no, just like everyone else in the class had and then he was like, \textit{What? so all of you are just in college for the social aspect?}...like suggesting that maybe like they're only going to school...because they want like the image or that they have ulterior motives or that they're not really passionate, hardworking scientists, which they absolutely are."
This accusation by the instructor may reveal biases or negative stereotypes about women, e.g., that women are not as hard working.

A few women in Study 2 reported that they had supportive instructors. For example, one senior student describes how her quantum professor made her feel supported.
She said that her quantum professor does ``...an excellent job at like, seeing like, what the students want and like making sure that they're understanding what he’s saying, so I feel like really supported in that environment, especially like [when I] go to office hours, and they are encouraging.''
This same student also describes another instructor in a similar way, thus revealing that there are a few instructors that are perceived positively by female students.

Women in Study 3 had more positive perceptions about their instructors than in Study 2 but some students had mixed experiences. For example, one student described her lecture instructor intervening when she was working with a condescending lab partner, but her lab instructor did not.
She recalled: ``The instructor for my class actually did notice that the student was being very difficult to work with and that he wasn't collaborating with the group, and she I think spoke to him about it." 
She explains that her lab instructor did not notice because: ``...a lot of our experiments were done outside of the physical classroom because, again with COVID, it was hard to sort of build those relationships with professors, so I think that was a part of the reason, but I also think another reason is when you're submitting work on time and when things are looking fine, a lot of professors don't go out of their way to look at the problems and a lot of professors, when necessary, will assume [things are fine] or [don't] want to have to deal with anything like that." 
From this student's quote, we get a sense that some faculty may not go out of their way to check in on students, thus leaving them unaware of any issues students face, especially in their interactions with each other.

Some students in Study 3 said that they feel comfortable around their instructors and perceived them as helpful. For example, one student explained that she doesn't attend office hours, unless she has a specific question. However, when she has gone, she has had positive experiences. She said attending office hours, ``was definitely helpful.'' She went after taking an exam. She explains: ``I had a question [about] something I did wrong, so I went and she helped me explain why it was wrong and then guided me through the correct answer, I think it was definitely helpful." 
Thus, we also see some variations in the experiences of women in this study. Some instructors are helpful but some are unaware of student dynamics.

\subsection{Cultural Domain}

In Study 1, Johnson reported the culture of students working together in a supportive environment. She also noted that from the focus group with students it was clear that students loved the physics building. For example, Johnson asked the focus group: \textit{What characterizes majoring in physics?}. Some of their responses were:
``The physics building."
``We live here",
``If I wasn't a commuter I'd be there 24-7",
``Spending more hours in the physics building outside class than you do sleeping",
``Is there life beyond this building? That's the question."
Thus, there is something about the physical space itself that creates this positive physics culture that student love to be around.

In faculty interviews, Johnson asked: \textit{What does it mean for someone to be a good physics student?}.
One faculty member said: ``to be curious and work hard. Curious in that you ask questions – you have to have enough confidence to ask questions, and realize – it does take some confidence to realize you can ask questions and it's not a bad reflection of you.''
Another faculty member said: ``We want everyone to be good physics students, but they don't have to all be great physics students. They have to be successful at acquiring various useful skills. They're not all going to be physicists, and we want them to be productive and happy."
These quotes from faculty reveal that they believe there are different ways to be a good physics student but in general think they should be able to ask questions, be resourceful, and be productive and happy. Faculty do not expect students to end up as professional physicists, but do expect them to develop important skills such as working hard and collaborating with others.

In Study 2, we get a strong sense of what the physics culture is like in this department not only from student and faculty interactions, but also from how instructors teach their courses. For example, one senior student explains that in her introductory physics courses, her instructors use disparaging language such as: 
\textit{This is trivial, and you should know this, right?}
She added: ``I felt like the...disrespectful behaviors that compose the culture and physics were taught in my first year at [my college], through like, professors using this language in their lectures that other people started to pick up on from their use."
From this account, we get a sense of how certain negative behaviors can be disseminated by professors in the classroom.

In Study 3, we earlier illustrated through student interactions that the physics culture has several masculine elements such as students being condescending towards their female peers, or examples of mansplaining. For example, one second year student shares that despite many of her male peers being helpful, some of them are condescending:
``A lot of the guys in my class are so helpful and so willing to help and don't approach things in a condescending manner at all, but then there are others who act like they know it all...I don't know if they do it because I'm a woman, but sometimes it feels like that, where they're like mansplaining so often in and outside the classroom, there is definitely a culture of mansplaining."
Here we see this student having mixed experiences with her peers, showing that some of them are ``so willing to help". However, there are still some students who mansplain. From her quote, it also seems like there is evidence of a culture of mansplaining not only inside the classroom but also outside.

Several women in Study 3 describe opportunities to work collaboratively with their peers and some show preference for working alone. However, it is concerning when students who are marginalized due to their identity (like women, women of color, etc.) are excluded from these collaborative opportunities. For example, we get some insight from one woman of color who feels alienated by her peers.
She explains that during group work, her peers ``...will kind of look at me and if we have to engage in sort of group work, there's usually not a lot of listening happening on their behalf. I think there's like this assumption that I'm, because I am a woman of color in STEM, I don't necessarily have the same background or knowledge as they [do], so a lot of the times I'll bring [up] points and it'll just be ignored and that’s something that I've had to deal with a lot."
Thus, it may be part of the physics culture in this department that other students do not take women of color or students of color seriously which may further exclude them in group settings. These behaviors of exclusion may be due to biases against women of color.

\subsection{Structural Domain}

In Study 1, we note that students report that their classes are interactive and incorporate group work.
For example, one student said: ``Most of the teachers will lecture for a half hour, 45 minutes, and then at some point in there ends up being some partner work or someone goes up to the board and works through a problem. So it's very interactive, instead of just being talked at. It's more a conversation with everybody in the class and the professor than information being thrown at you, because that’s not helpful. I think that's a big thing to me as to why I enjoy it and I have learned so much from it."
We see from this quote that the student actually enjoys the interactive elements that are structured into class and finds them to be useful in their learning.

Because Johnson interviewed faculty members, she was able to get their perspectives as to how physics classes are structured.
Faculty emphasized that group work is built into their courses and that
``lectures are very strongly de-emphasized in favor of a lot of group work, a lot of interaction between faculty and students, fairly short lab exercises which are pertinent to the material being taught rather than having separate lab sessions which could be a week behind or a week ahead of the class material at the time."
Johnson reported that the physics faculty make use of physics education research through ``constant department-wide use of both formative and summative assessment." Thus, this department strongly emphasizes teaching interactively which is backed by research.

In Study 2, most faculty choose to teach in a traditional style, mainly lecturing, while few faculty provide opportunities for collaborative work, such group problem solving or clicker questions. There are also few instructors who incorporate active learning (i.e., flipped classrooms). 
We confirmed this information with instructors at this institution.

In Study 3, many of the women described opportunities for group work during class. We note that importance of group work came up not only in response to questions about peer interactions, but also in their suggestions for instructors. Specifically, some women called for their instructors to encourage and incorporate more peer collaboration into the class structure.
For example, one student suggested that instructors should change how in-class groups are formed by: 
``assigning people to different groups and having rotations where people aren't always stuck with the same group but also have enough time working with other people to break down these barriers..."
This student argued that implementing rotations would allow students to get to know many people in class and can introduce them to students with different backgrounds, and thus ``break down these barriers" that may arise  due to differences in background or identity.

\subsection{Disciplinary Domain}

In Study 1, Johnson reported many instances where faculty were ``reprimanding students who failed to work equitably in groups.'' It is important to note that faculty in this physics department value that everyone learns during group work, as opposed to working efficiently. One faculty member recalled working with a student who was dominating group work during a lab. She said that this student was controlling all the materials, so she told him that he had to let other people have a chance, at which point he backed up and stood far away from his group. She told him he did not have to stand so far away but added, ``You can’t only participate when you're building, that's not OK. It can't be \textit{I'm either in charge or I'm out of here, guys.}" 
This is an explicit example of faculty members taking action when students are not working according to set norms which the faculty member modeled.

In Study 2, we note the lack of disciplinary actions from faculty through peer and instructor interactions, described by the women. However, the women also explicitly call this out in their suggestions for faculty members. One student suggested that instructors should be responsible for establishing and maintaining a positive learning environment.
She said, ``I want [instructors] to be obligated, when they witness what happens in their classes with other students, to confront them, the students who [engage in microaggressions] because the problem just isn't with professors, it's with professors and students-students feel like they can do that-these things after they see their professors do it."
This student suggested that instructors should be responsible not only for their actions, but also to confront students who commit microaggressions in their classrooms.

Through faculty interactions and class structure, some women from Study 3 feel that faculty members are not aware of issues of misconduct and therefore cannot take disciplinary action. One of the women described how faculty members seem surprised if a student speaks up about not feeling represented in the learning space.
She said that professors were ``surprised by students speaking up and by students feeling this way, and so I guess after that, they had a conversation and the next class time for example [professors will] try to make it known to students \textit{Hey, we should be engaging with each other}, or trying to discretely say...there should be more interactions with individuals that aren't necessarily White or that are women in STEM and so I think that's been extent [of] things usually, when a student speaks up about it I've seen temporary discussions of it with professors and maybe in the classroom setting but it feels almost as if, a few weeks later it kind of goes back to what it used to be." 
This quote suggests that when faculty take any disciplinary action (if ever), the effects are very temporary and students revert to their old behaviors. Also, it seems like this disciplining may only occur when students complain about something and it is not reinforced.

\section{Discussion}
Based on the example quotes from women in each study, we get a sense of what each physics department culture is like from the perspective of undergraduate women (supplemented by faculty interviews in Study 1), despite there being limited information. Consistent with Standpoint theory, 
comparing undergraduate women's voices in this comparative case study helps us to understand how they  experience the overall physics culture in their departments so that other physics departments can contemplate how to make their learning environments equitable.
Furthermore, utilizing all the three frameworks emphasizes that the voices of underrepresented groups such as undergraduate women in physics should be highlighted and that classroom instructors 
and faculty members in general 
have power in several domains to create safe and equitable learning environments. With that being said, we use for broader comparison the table provided in Johnson's study, Table \ref{table:4.5}, which describes the physics department in Study 1 in the context of the Domains of Power compared to a prototypical physics department. This table can be valuable for comparing the three studies
(see Table 
\ref{table:4.1}). It is also important to emphasize that the four Domains of Power interact with one another so the quotes discussed in one domain may inform the physics culture in multiple domains.

\begin{table*}[tb]
\small
\begin{center}
\begin{minipage}{0.95\textwidth}
    \caption{\raggedright This table is adapted from Johnson's study \cite{johnson2020intersectional}. Comparing characteristics of Domains of Power in a prototypical physics department and the department under study in Study 1}\label{table:4.5}
    \begin{ruledtabular}
    \begin{tabular}{lll}
    \toprule
    Domain & Prototypical physics departments & Physics department from Study 1 \\ \hline
    \multirow{3}{5em}{Interpersonal} & \multirow{3}{18em}{Women of color experience isolation and experience microaggressions.} & \multirow{3}{26em}{Students are friendly and helpful; they work on problems together (even if they don't really like group work) and socialize together.} \\
    & & \\
    & & \\
    \multirow{3}{5em}{Cultural} & \multirow{3}{18em}{Physics is a competition; good physicists are natural geniuses who work in isolation.} & \multirow{3}{26em}{Physics is collaborative; success in physics results from hard work and practice; physicists can be wrong.} \\
    & & \\
    & & \\
    \multirow{3}{5em}{Structural} & \multirow{3}{18em}{Classes consist of faculty lecturing; high levels of student misconception and failure are expected.} & \multirow{3}{26em}{Faculty and students are highly interactive; students work collaboratively during class to learn physics and solve problems.} \\
    & & \\
    & & \\
    \multirow{3}{5em}{Disciplinary} & \multirow{3}{18em}{Faculty do not intervene in student-student interactions.} & \multirow{3}{26em}{When students contest this collaborative culture and attempt to assert a prototypical physics identity faculty intervene.}\\
    & & \\
    & & \\
    \end{tabular}
    \end{ruledtabular}
\end{minipage}
\end{center}
\end{table*}

\subsection{Interpersonal domain}

Johnson's study illustrates a physics department where students are willing to help each other. Students in her study perceived other students and instructors as welcoming. They could go to others when they struggle (also hinting at the physics cultural domain). From faculty interviews, it seems like faculty go out of their way to be available so students can ask questions in addition to encouraging questions from students (also hinting at the supportive physics cultural domain). Johnson emphasized that the interpersonal domain described in her study was opposite of the one in a prototypical physics department, see Table \ref{table:4.5}. 

On the other hand, based on the interpersonal domain pertaining to student interactions in Studies 2 and 3, it appears that those departments' culture emulates a prototypical physics department to different degrees, where some students are often excluded from meaningful collaboration by those from majority groups (e.g., male students). 
We note that grounded in Standpoint theory, which emphasizes the points of view of traditionally marginalized groups such as the undergraduate women,  even without interviews with physics faculty members, Studies 2 and 3 provide a reasonably good representation of the interpersonal domain for women physics majors.
Although Study 3 described a department with a better physics culture than Study 2, in Study 3, a woman of color described being isolated to the point where she would not work with White peers regardless of gender because she felt discriminated due to her racial identity. We also note that in Study 3, there were many accounts of men mansplaining and being condescending to women (again hinting at the physics cultural domain).

\subsection{Cultural domain}

In regard to the cultural domain, Study 1 reveals a collaborative environment, created and enforced by faculty members. Students chose to collaborate and recognize its value despite describing themselves as antisocial. There is a culture in this department where students can collaborate without any anxiety. We also see evidence of a culture within the physical space of the physics building itself, where students in the focus group described spending a lot of time inside doing physics. In regard to the social culture, faculty believed that being a good physics student means being curious and hard working. Faculty and students both acknowledge that students can be wrong, in fact, it is anticipated while learning physics. Even students recognized that faculty won't ``frown'' at them if they are confused. Faculty members also acknowledged that not every physics student would become a physicist, so they encouraged other pathways besides applying to a Ph.D. program. This physics culture may put less pressure on students to pursue career options that do not include academia or careers as professional physicists in national labs or physics-related industries. This rejection by faculty of a traditional image of a physicist by accepting and promoting alternative ways to be a physics person can potentially boost students' sense of belonging. Students in this department can be accepted for being themselves even if they choose to not pursue a physics Ph.D.

The physics departments described in Studies 2 and 3 seem to reflect more prototypical physics culture (especially the one in Study 2). For example, there are many instances of mansplaining and students being treated like they should know certain physics concepts. Thus, such departments may foster a culture that tells students that they need to be geniuses to succeed in physics and that good physicists simply do not struggle. Although some faculty members from Studies 2 and 3 (especially in Study 3) seemed very welcoming to questions, some of these undergraduate women students still felt that they could not seek out help.  
Women students from Study 2 seem to isolate themselves from men.  This can again be due to the physics culture fostering and reinforcing mansplaining and encouraging men to be condescending towards their female peers. In Study 3, the woman of color, Paulina, described herself as an island who did not want to collaborate even with women.

\subsection{Structural domain}
The structural domain (e.g., class structure) of Study 1 appeared to be very interactive and collaborative. Because group work was incorporated into the class structure, students had many opportunities to work with their peers. This itself can positively impact the cultural domain, allowing students to feel safe amongst their peers to ask questions without the fear of being wrong or judged. It also seems like all faculty make an effort to utilize surveys and assessments that are validated and backed by research.

Study 1 contrasts to the other two studies, especially Study 2, in which courses were taught more traditionally. In Study 2, there are many faculty members in the physics department so it is difficult to have a standardized method for teaching. There is some evidence from interviews of active learning elements such as group work and the use of clicker questions during lectures. However, this is the choice of individual instructors. In Study 3, there is more evidence of collaborative work implemented by instructors. Based on peer interactions, it appears that some instructors incorporate group work. However, like Study 2, it seems to be more of the instructors' choice as opposed to a standardized or common practice, according to students' suggestions for their instructors.

\subsection{Disciplinary domain}

We emphasize the stark differences in the disciplinary domain between the studies discussed here. Study 1 has direct evidence of faculty disciplining students due to failure to work equitably in groups. According to Johnson, male faculty take responsibility for gender issues \cite{johnson2020intersectional}. Thus, women faculty do not bear the burden or responsibility for addressing gender issues, such as sexism. This physics department culture also corresponds to one in which the women physics students believe their faculty would protect them from negative interactions with peers. It is unclear how consistently faculty members have to address disciplinary issues that arise repetitively but Johnson emphasizes that in order to construct a collaborative structural domain (class structure), ``it requires constant maintenance.''

In Study 2, there seems to be a severe lack of disciplining from instructors. This was a major issue voiced by women in study 2, and they suggested that instructors need to reprimand students for misconduct. It could be the that the physics cultural domain is so prototypical and negative behaviors are reinforced so much that disciplinary action seems taboo or against the departmental culture itself. For example, in Study 2, women often felt marginalized by their own instructors, and some felt that the instructors are responsible for modeling negative behaviors towards women. Thus, they were less likely to recognize and discipline such behaviors when male students displayed them.

Study 3 is different from both Studies 1 and 2 in that the women commented that their faculty seemed oblivious of issues between students even though they themselves were helpful and supportive. 
This type of unawareness of negative interactions between students with different identities suggests that many instructors are not ``primed" or prepared to recognize issues in their classroom. As one woman said, ``if they're not aware of [issues]...there's no way for them to personally intervene...'' Thus, there needs to be more awareness before faculty can take action. It is also interesting that some students noted that if faculty are made aware of issues in their classrooms, they address it superficially so that this does not have a lasting effect. This lack of lasting effect can be evidence for a physics culture that reinforces and models the prototypical image of a physicist.

\section{Conclusions}
In this paper, we carried out a comparative case study
primarily using the Domains of Power framework but also drawing inspiration from other synergistic frameworks.
All of these studies use qualitative methods to investigate undergraduate women's experiences in physics and astronomy departments and how their accounts portray different types of physics cultures. The first study is Johnson's study in a small physics department at a small predominantly White liberal arts college. Her study illustrated an overwhelmingly supportive and positive physics learning environment where students work together and faculty are always encouraging and supportive. The second is Santana and Singh's study from a large predominantly White research institution. This study revealed a very unwelcoming physics learning environment where many male students and male faculty members negatively impacted the experiences of women students. Lastly, the third study is Santana and Singh's more recent study in a mid-sized physics department at a small private predominantly White liberal arts college. This study highlighted how intersectional issues in identity (e.g., being traditionally marginalized both by gender and race) can play a negative role in students' experiences in an undergraduate physics program.

The three frameworks used in these studies can be used as guidelines for how physics instructors can approach interactions with students, structure the classrooms, work with colleagues, etc.
Based upon the Domains of Power and HELPIEE frameworks, we emphasize that physics instructors have a lot of power not only in their classrooms (structural domain), but also in the physics cultural domain to empower students.
Standpoint theory suggests that it is important for physics faculty to listen to women's experiences in physics in order to address inequities. 
It is important for faculty to utilize this power not only inside the classroom, but also outside the classroom in order to continue supporting students by modeling alternative ways to be good physics students, as in Study 1.

We can use the physics department described in Johnson's study as a model to transform prototypical physics departments to be inclusive and equitable for even the most marginalized groups.
Considering the fact that the physics department in Study 1 only had a handful of faculty members, the physics culture there
may appear to be an unrealistic standard for all physics departments to achieve.  However, although having only a handful of faculty may make it easier for them to communicate and come to a consensus on shared norms to make the physics culture similar to that depicted in Study 1, all physics departments regardless of the size of their faculty and student body should strive to establish similar culture. We saw that students formed a community with each other and felt like they were ``all in it together,'', with the mindset of ``why wouldn't they help me?'' \cite{johnson2020intersectional}, which illustrates an opposite culture to the infamous prototypical competitive and isolated physics culture.
Furthermore, certain elements of Study 1 such as having faculty be on the same page regarding how to identify disruptive and condescending student behavior and disciplining students as necessary, as well as having class structure supportive of all students etc. can be incorporated in physics departments of any size. As a faculty member in Johnson's study said, ``You can't only participate when you're building, that's not OK'' \cite{johnson2020intersectional}. 

For example, there are about 50 physics faculty in the physics department in Study 2. This may lead to fragmentation in this department making it more difficult to dismantle a toxic physics culture or to successfully restructure classes.=
However, using this as an excuse to not take action is unacceptable, and the physics department cannot be complacent about equity and inclusion.
Also, having a critical number of committed faculty members dedicated towards reforming their department culture is essential for creating an inclusive and equitable physics environment. 
We emphasize that even in Study 2, there were some faculty members who supported the women students. Thus, some faculty members who are not actively contributing to a toxic physics environment can work together to catalyze change and
with enough faculty support and resources, there can be a sustained and systemic positive change in any sized physics department.
In addition, it would be interesting to investigate the cultural and pedagogical differences of these three physics departments to ascertain if they attract faculty with less typical views of what a physicist is and approaches to teaching.

We note that in Studies 2 and 3, the interviewer asked the undergraduate women about what suggestions they might have to improve their own experiences. 
These suggestions are summarized in both papers.
We emphasize that many of these suggestions (like those for instructors), do not require many resources, but require desire to change, time, and effort.
We strongly encourage physics faculty to consider these suggestions and brainstorm with others in their department to begin to make positive changes as may be more appropriate for a particular physics department.
Systemic changes would require more resources in addition to time and effort. 
However, physics departments should not view these as impossible changes to implement because they are achievable when a critical number of faculty members do take part in this mission.
In conclusion, it is the responsibility of physics instructors and faculty to use their power in all the four Domains of Power in order to fully transform a physics culture including inside and outside the classrooms and in student-student and student-faculty interactions.

\section{Acknowledgements}
We are very grateful to Dr. Robert P. Devaty for reviewing and providing feedback on this manuscript.

\bibliography{ref}

\end{document}